\input harvmac
\input epsf

\def\figin{\epsfcheck\figin}\def\figins{\epsfcheck\figins}
\def\epsfcheck{\ifx\epsfbox\UnDeFiNeD
\message{(NO epsf.tex, FIGURES WILL BE IGNORED)}
\gdef\figin##1{\vskip2in}\gdef\figins##1{\hskip.5in}% blank space instead
\else\message{(FIGURES WILL BE INCLUDED)}%
\gdef\figin##1{##1}\gdef\figins##1{##1}\fi}
\def\DefWarn#1{}
\def\figinsert{\goodbreak\topinsert}
\def\ifig#1#2#3#4{\DefWarn#1\xdef#1{fig.~\the\figno}
\writedef{#1\leftbracket fig.\noexpand~\the\figno}%
\figinsert\figin{\centerline{\epsfxsize=#3mm \epsfbox{#2}}}
\bigskip\medskip\centerline{\vbox{\baselineskip12pt
\advance\hsize by -1truein\noindent\footnotefont{\sl Fig.~\the\figno:}\sl\ #4}}
\bigskip\endinsert\noindent\global\advance\figno by1}

\def\Z{{\bf Z}}

\def\a{\alpha}
\def\b{\beta}

\def\e{\epsilon}

\def\th{\theta}

\def\l{\lambda}
\def\L{\Lambda}
\def\m{\mu}

\def\s{\sigma}
\def\t{\tau}

\def\F{\Phi}
\def\w{\omega}

\def\v{\varphi}

\def\d{\partial}
\def\dbar{{\overline\partial}}

\def\Tr{{\rm Tr}}
\def\hf{{1\over 2}}
\def\cO{{\cal O}}

\def\cF{{\cal F}}
\def\cN{{\cal N}}
\def\cW{{\cal W}}

\def\({\bigl(}
\def\){\bigr)}
\def\<{\langle\,}
\def\>{\,\rangle}

\def\D{\Delta}
\def\eps{\epsilon}
\def\]{\right]}
\def\[{\left[}

\def\tN{{\widetilde{N}}}

\lref\thooft{
G.~'t Hooft, ``A Planar Diagram Theory For Strong Interactions,''
Nucl.\ Phys.\ B {\bf 72}, 461 (1974).}

\lref\mm{
P.~Ginsparg and G.~W.~Moore,
``Lectures On 2-D Gravity And 2-D String Theory,''
arXiv:hep-th/9304011.}

\lref\mmm{
P.~Di Francesco, P.~Ginsparg and J.~Zinn-Justin,
``2-D Gravity and random matrices,''
Phys.\ Rept.\  {\bf 254}, 1 (1995)
[arXiv:hep-th/9306153]}

\lref\kontsevich{M.~Kontsevich,
``Intersection Theory On The Moduli Space Of Curves And The Matrix
Airy Function,'' Commun.\ Math.\ Phys.\ {\bf 147}, 1 (1992).}

\lref\wittentop{E.~Witten,
``On The Structure Of The Topological Phase Of Two-Dimensional
Gravity,'' Nucl.\ Phys.\ B {\bf 340}, 281 (1990)}

\lref\bfss{T.~Banks, W.~Fischler, S.~H.~Shenker and L.~Susskind,
``M theory as a matrix model: A conjecture,'' Phys.\ Rev.\ D {\bf 55},
5112 (1997) [arXiv:hep-th/9610043].}

\lref\adscft{
O.~Aharony, S.~S.~Gubser, J.~M.~Maldacena, H.~Ooguri and Y.~Oz,
``Large $N$ field theories, string theory and gravity,'' Phys.\ Rept.\
{\bf 323}, 183 (2000) [arXiv:hep-th/9905111].  }

\lref\ghv{D. Ghoshal and C. Vafa, ``$c=1$ string as the topological theory
of the conifold,'' Nucl.\ Phys.\ B {\bf 453}, 121 (1995)
[arXiv:hep-th/9506122].}

\lref\klmkv{S. Kachru, A. Klemm, W. Lerche, P. Mayr, C. Vafa,
``Nonperturbative Results on the Point Particle Limit of $N=2$
Heterotic String Compactifications,'' Nucl.\ Phys.\ B {\bf 459}, 537
(1996) [arXiv:hep-th/9508155].}

\lref\klw{A. Klemm, W. Lerche, P. Mayr, C.Vafa, N. Warner,
``Self-Dual Strings and $N=2$ Supersymmetric Field Theory,''
 Nucl.\ Phys. \ B {\bf 477}, 746 (1996)
[arXiv:hep-th/9604034].}

\lref\kmv{S. Katz, P. Mayr, C. Vafa,
``Mirror symmetry and Exact Solution of 4D $N=2$ Gauge Theories I,''
Adv.\ Theor.\ Math.\ Phys. {\bf 1}, 53 (1998)
[arXiv:hep-th/9706110].}

\lref\gv{R.~Gopakumar and C.~Vafa,
``On the gauge theory/geometry correspondence,'' Adv.\ Theor.\ Math.\
Phys.\ {\bf 3}, 1415 (1999) [arXiv:hep-th/9811131].  }

\lref\edel{J.D. Edelstein, K. Oh and R. Tatar, ``Orientifold,
geometric transition and large $N$ duality for $SO/Sp$ gauge
theories,'' JHEP {\bf 0105}, 009 (2001) [arXiv:hep-th/0104037].}

\lref\dasg{K. Dasgupta, K. Oh and R. Tatar, ``Geometric transition, large
$N$ dualities and MQCD dynamics,'' Nucl. Phys.  B {\bf 610}, 331
(2001) [arXiv:hep-th/0105066]\semi -----, ``Open/closed string
dualities and Seiberg duality from geometric transitions in
M-theory,'' [arXiv:hep-th/0106040]\semi -----, ``Geometric transition
versus cascading solution,'' JHEP {\bf 0201}, 031 (2002)
[arXiv:hep-th/0110050].}

\lref\hv{K. Hori and C. Vafa,
``Mirror Symmetry,'' [arXiv:hep-th/0002222].}

\lref\hiv{K. Hori, A. Iqbal and C. Vafa,
``D-Branes And Mirror Symmetry,'' [arXiv:hep-th/0005247].}

\lref\vaug{C.~Vafa,
``Superstrings and topological strings at large $N$,''
J.\ Math.\ Phys.\  {\bf 42}, 2798 (2001)
[arXiv:hep-th/0008142].}

\lref\civ{
F.~Cachazo, K.~A.~Intriligator and C.~Vafa,
``A large $N$ duality via a geometric transition,''
Nucl.\ Phys.\ B {\bf 603}, 3 (2001)
[arXiv:hep-th/0103067].}

\lref\ckv{
F.~Cachazo, S.~Katz and C.~Vafa,
``Geometric transitions and $N = 1$ quiver theories,''
arXiv:hep-th/0108120.}

\lref\cfikv{
F.~Cachazo, B.~Fiol, K.~A.~Intriligator, S.~Katz and C.~Vafa, ``A
geometric unification of dualities,'' Nucl.\ Phys.\ B {\bf 628}, 3
(2002) [arXiv:hep-th/0110028].}

\lref\cv{
F.~Cachazo and C.~Vafa, ``$N=1$ and $N=2$ geometry from fluxes,''
arXiv:hep-th/0206017.}

\lref\ov{
H.~Ooguri and C.~Vafa, ``Worldsheet derivation of a large $N$ duality,''
Nucl.\ Phys.\ B {\bf 641} (2002) 3, [arXiv:hep-th/0205297].}

\lref\av{
M.~Aganagic and C.~Vafa, ``$G_2$ manifolds, mirror symmetry, and
geometric engineering,'' arXiv:hep-th/0110171.}

\lref\digra{
D.~E.~Diaconescu, B.~Florea and A.~Grassi, ``Geometric transitions and
open string instantons,'' arXiv:hep-th/0205234.}

\lref\amv{
M.~Aganagic, M.~Marino and C.~Vafa, ``All loop topological string
amplitudes from Chern-Simons theory,'' arXiv:hep-th/0206164.}

\lref\dfg{
D.~E.~Diaconescu, B.~Florea and A.~Grassi, ``Geometric transitions,
del Pezzo surfaces and open string instantons,''
arXiv:hep-th/0206163.}

\lref\kkl{
S.~Kachru, S.~Katz, A.~E.~Lawrence and J.~McGreevy, ``Open string
instantons and superpotentials,'' Phys.\ Rev.\ D {\bf 62}, 026001
(2000) [arXiv:hep-th/9912151].}

\lref\bcov{
M.~Bershadsky, S.~Cecotti, H.~Ooguri and C.~Vafa, ``Kodaira-Spencer
theory of gravity and exact results for quantum string amplitudes,''
Commun.\ Math.\ Phys.\ {\bf 165}, 311 (1994) [arXiv:hep-th/9309140].
}

\lref\witcs{
E.~Witten, ``Chern-Simons gauge theory as a string theory,''
arXiv:hep-th/9207094.  }

\lref\naret{I. Antoniadis, E. Gava, K.S. Narain, T.R. Taylor,
``Topological Amplitudes in String Theory,'' Nucl.\ Phys.\ B\ {\bf
413}, 162 (1994) [arXiv:hep-th/9307158].}

\lref\witf{E. Witten,
``Solutions Of Four-Dimensional Field Theories Via M Theory,'' Nucl.\
Phys.\ B\ {\bf 500}, 3 (1997) [arXiv:hep-th/9703166].}

\lref\shenker{
S.~H.~Shenker, ``The Strength Of Nonperturbative Effects In String
Theory,'' in Proceedings Cargese 1990, {\it Random surfaces and
quantum gravity}, 191--200.  }

\lref\berwa{
M.~Bershadsky, W.~Lerche, D.~Nemeschansky and N.~P.~Warner, ``Extended
$N=2$ superconformal structure of gravity and W gravity coupled to
matter,'' Nucl.\ Phys.\ B {\bf 401}, 304 (1993)
[arXiv:hep-th/9211040]}

\lref\akemann{
G.~Akemann, ``Higher genus correlators for the Hermitian matrix model
with multiple cuts,'' Nucl.\ Phys.\ B {\bf 482}, 403 (1996)
[arXiv:hep-th/9606004].  }

\lref\wiegmann{P.B. Wiegmann and A. Zabrodin, ``Conformal maps
and integrable hierarchies,'' arXiv:hep-th/9909147.}

\lref\cone{R.~Dijkgraaf and C.~Vafa, to appear.}

\lref\kazakov{
S.~Y.~Alexandrov, V.~A.~Kazakov and I.~K.~Kostov, ``Time-dependent
backgrounds of 2D string theory,'' arXiv:hep-th/0205079.}

\lref\dj{
S.~R.~Das and A.~Jevicki, ``String Field Theory And Physical
Interpretation Of $D=1$ Strings,'' Mod.\ Phys.\ Lett.\ A {\bf 5}, 1639
(1990).}

\lref\givental{
A.B.~Givental, ``Gromov-Witten invariants and quantization of
 quadratic hamiltonians,'' arXiv:math.AG/0108100.}

\lref\op{
A.~Okounkov and R.~Pandharipande, ``Gromov-Witten theory, Hurwitz
theory, and completed cycle,'' arXiv:math.AG/0204305.}

\lref\dijk{ R.~Dijkgraaf, ``Intersection theory, integrable hierarchies and
topological field theory,'' in Cargese Summer School on {\it New
Symmetry Principles in Quantum Field Theory} 1991,
[arXiv:hep-th/9201003].}

\lref\gw{
D.~J.~Gross and E.~Witten, ``Possible Third Order Phase Transition In
The Large $N$ Lattice Gauge Theory,'' Phys.\ Rev.\ D {\bf 21}, 446
(1980).}

\lref\sw{
N.~Seiberg and E.~Witten, ``Electric-magnetic duality, monopole
condensation, and confinement in $N=2$ supersymmetric Yang-Mills
theory,'' Nucl.\ Phys.\ B {\bf 426}, 19 (1994) [Erratum-ibid.\ B {\bf
430}, 485 (1994)] [arXiv:hep-th/9407087].}

\lref\kkv{
S.~Katz, A.~Klemm and C.~Vafa, ``Geometric engineering of quantum
field theories,'' Nucl.\ Phys.\ B {\bf 497}, 173 (1997)
[arXiv:hep-th/9609239].}

\lref\taylor{
W.~I.~Taylor, ``D-brane field theory on compact spaces,'' Phys.\
Lett.\ B {\bf 394}, 283 (1997) [arXiv:hep-th/9611042].}

\lref\kmmms{
S.~Kharchev, A.~Marshakov, A.~Mironov, A.~Morozov and S.~Pakuliak,
``Conformal matrix models as an alternative to conventional
multimatrix models,'' Nucl.\ Phys.\ B {\bf 404}, 717 (1993)
[arXiv:hep-th/9208044].}

\lref\dv{
R.~Dijkgraaf and C.~Vafa, ``Matrix models, topological strings, and
supersymmetric gauge theories,'' arXiv:hep-th/0206255.}

\lref\kostov{
I.~K.~Kostov, ``Gauge invariant matrix model for the A-D-E closed
strings,'' Phys.\ Lett.\ B {\bf 297}, 74 (1992)
[arXiv:hep-th/9208053].  }

\lref\ot{
K.~h.~Oh and R.~Tatar, ``Duality and confinement in $N=1$
supersymmetric theories from geometric transitions,''
arXiv:hep-th/0112040.}

\lref\ovknot{
H.~Ooguri and C.~Vafa, ``Knot invariants and topological strings,''
Nucl.\ Phys.\ B {\bf 577}, 419 (2000) [arXiv:hep-th/9912123].  }

\lref\dvv{
R.~Dijkgraaf, H.~Verlinde and E.~Verlinde, ``Loop Equations And
Virasoro Constraints In Nonperturbative 2-D Quantum Gravity,'' Nucl.\
Phys.\ B {\bf 348}, 435 (1991).}

\lref\kawai{
M.~Fukuma, H.~Kawai and R.~Nakayama, ``Continuum Schwinger-Dyson
Equations And Universal Structures In Two-Dimensional Quantum
Gravity,'' Int.\ J.\ Mod.\ Phys.\ A {\bf 6}, 1385 (1991).}

\lref\nekrasov{
N.~A.~Nekrasov, ``Seiberg-Witten prepotential from instanton
counting,'' arXiv:hep-th/0206161.}

\lref\ki{
I.~K.~Kostov, ``Bilinear functional equations in 2D quantum gravity,''
in Razlog 1995, {\it New trends in quantum field theory}, 77--90,
[arXiv:hep-th/9602117].}

\lref\kii{
I.~K.~Kostov, ``Conformal field theory techniques in random matrix
models,'' arXiv:hep-th/9907060.}

\lref\morozov{
A.~Morozov, ``Integrability And Matrix Models,'' Phys.\ Usp.\ {\bf
37}, 1 (1994) [arXiv:hep-th/9303139].}

\lref\fo{
H.~Fuji and Y.~Ookouchi, ``Confining phase superpotentials for SO/Sp
gauge theories via geometric transition,'' arXiv:hep-th/0205301.  }

\lref\marinor{M. Marino, ``Chern-Simons theory, matrix integrals, and
perturbative three-manifold invariants,'' [arXiv:hep-th/0207096].}

\lref\agnt{
I.~Antoniadis, E.~Gava, K.~S.~Narain and T.~R.~Taylor, ``Topological
amplitudes in string theory,'' Nucl.\ Phys.\ B {\bf 413}, 162 (1994)
[arXiv:hep-th/9307158].}

\lref\vw{
C.~Vafa and E.~Witten, ``A Strong coupling test of S duality,'' Nucl.\
Phys.\ B {\bf 431}, 3 (1994) [arXiv:hep-th/9408074].}

\lref\ps{
J.~Polchinski and M.~J.~Strassler, ``The string dual of a confining
four-dimensional gauge theory,'' arXiv:hep-th/0003136.}

\lref\kkn{
V.~A.~Kazakov, I.~K.~Kostov and N.~A.~Nekrasov, ``D-particles, matrix
integrals and KP hierarchy,'' Nucl.\ Phys.\ B {\bf 557}, 413 (1999)
[arXiv:hep-th/9810035].}

\lref\kpw{
A.~Khavaev, K.~Pilch and N.~P.~Warner, ``New vacua of gauged $N = 8$
supergravity in five dimensions,'' Phys.\ Lett.\ B {\bf 487}, 14
(2000) [arXiv:hep-th/9812035].}

\lref\ks{
S.~Kachru and E.~Silverstein, ``4d conformal theories and strings on
orbifolds,'' Phys.\ Rev.\ Lett.\ {\bf 80}, 4855 (1998)
[arXiv:hep-th/9802183].  }

\lref\bj{
M.~Bershadsky and A.~Johansen, ``Large $N$ limit of orbifold field
theories,'' Nucl.\ Phys.\ B {\bf 536}, 141 (1998)
[arXiv:hep-th/9803249].}

\lref\lnv{
A.~E.~Lawrence, N.~Nekrasov and C.~Vafa, ``On conformal field theories
in four dimensions,'' Nucl.\ Phys.\ B {\bf 533}, 199 (1998)
[arXiv:hep-th/9803015].  }

\lref\bkv{
M.~Bershadsky, Z.~Kakushadze and C.~Vafa, ``String expansion as large
$N$ expansion of gauge theories,'' Nucl.\ Phys.\ B {\bf 523}, 59 (1998)
[arXiv:hep-th/9803076].}

\lref\dvii{
R.~Dijkgraaf and C.~Vafa, ``On geometry and matrix models,''
arXiv:hep-th/0207106.}

\lref\vy{
G.~Veneziano and S.~Yankielowicz, ``An Effective Lagrangian For The
Pure $N=1$ Supersymmetric Yang-Mills Theory,'' Phys.\ Lett.\ B {\bf
113}, 231 (1982).  }

\lref\sinhv{
S.~Sinha and C.~Vafa, ``$SO$ and $Sp$ Chern-Simons at large $N$,''
arXiv:hep-th/0012136.  }

\lref\aahv{B.~Acharya, M.~Aganagic, K.~Hori and C.~Vafa,
``Orientifolds, mirror symmetry and superpotentials,''
arXiv:hep-th/0202208.}

\lref\dorey{N. Dorey, ``An Elliptic Superpotential for Softly
Broken $N=4$ Supersymmetric Yang-Mills,'' JHEP {\bf 9907}, 021 (1999)
[arXiv:hep-th/9906011].}

\lref\ls{
R.~G.~Leigh and M.~J.~Strassler, ``Exactly marginal operators and
duality in four-dimensional $N=1$ supersymmetric gauge theory,'' Nucl.\
Phys.\ B {\bf 447}, 95 (1995) [arXiv:hep-th/9503121].
}

\lref\klyt{
A.~Klemm, W.~Lerche, S.~Yankielowicz and S.~Theisen,
``Simple singularities and $N=2$ supersymmetric Yang-Mills theory,''
Phys.\ Lett.\ B {\bf 344}, 169 (1995)
[arXiv:hep-th/9411048]}

\lref\af{
P.~C.~Argyres and A.~E.~Faraggi, ``The vacuum structure and spectrum
of $N=2$ supersymmetric $SU(n)$ gauge theory,'' Phys.\ Rev.\ Lett.\
{\bf 74}, 3931 (1995) [arXiv:hep-th/9411057].}

\lref\mo{C.~Montonen and D.~I.~Olive,
``Magnetic Monopoles As Gauge Particles?,'' Phys.\ Lett.\ B {\bf 72},
117 (1977).  }

\lref\gvw{S.~Gukov, C.~Vafa and E.~Witten,
``CFT's from Calabi-Yau four-folds,'' Nucl.\ Phys.\ B {\bf 584}, 69
(2000) [Erratum-ibid.\ B {\bf 608}, 477 (2001)]
[arXiv:hep-th/9906070].  }

\lref\tv{
T.~R.~Taylor and C.~Vafa, ``RR flux on Calabi-Yau and partial
supersymmetry breaking,'' Phys.\ Lett.\ B {\bf 474}, 130 (2000)
[arXiv:hep-th/9912152].  }

\lref\mayr{
P.~Mayr, ``On supersymmetry breaking in string theory and its
realization in brane worlds,'' Nucl.\ Phys.\ B {\bf 593}, 99 (2001)
[arXiv:hep-th/0003198].  }

\lref\qfvy{A.~de la Macorra and G.G.~Ross, Nucl.\ Phys.\ B {\bf 404},
321 (1993)\semi C. P. Burgess, J.-P. Derendinger, F. Quevedo,
M. Quiros, ``On Gaugino Condensation with Field-Dependent Gauge
Couplings,'' Annals Phys. {\bf 250}, 193 (1996)
[arXiv:hep-th/9505171].}

\lref\div{
A. D'Adda, A.C. Davis, P. Di Vecchia and P. Salomonson, Nucl. Phys. B
{\bf 222}, 45 (1983).}

\lref\berva{
N. Berkovits and C. Vafa, ``$N=4$ Topological Strings,'' Nucl. Phys. B
{\bf 433} 123 (1995) [arXiv:hep-th/9407190].}

\lref\novikov{
V.~A.~Novikov, M.~A.~Shifman, A.~I.~Vainshtein and V.~I.~Zakharov,
``Instanton Effects In Supersymmetric Theories,'' Nucl.\ Phys.\ B {\bf
229}, 407 (1983).  }

\lref\dk{
N.~Dorey and S.~P.~Kumar, ``Softly-broken $N = 4$ supersymmetry in the
large-$N$ limit,'' JHEP {\bf 0002}, 006 (2000) [arXiv:hep-th/0001103].
}

\lref\ds{
N.~Dorey and A.~Sinkovics, ``$N = 1^*$ vacua, fuzzy spheres and
integrable systems,'' JHEP {\bf 0207}, 032 (2002)
[arXiv:hep-th/0205151].  }

\lref\ads{
I.~Affleck, M.~Dine and N.~Seiberg, ``Supersymmetry Breaking By
Instantons,'' Phys.\ Rev.\ Lett.\ {\bf 51}, 1026 (1983).  }

\lref\bele{
D. Berenstein, V. Jejjala and R. G. Leigh, ``Marginal and Relevant
Deformations of $N=4$ Field Theories and Non-Commutative Moduli Spaces
of Vacua,'' Nucl.Phys. B {\bf 589} 196 (2000) [arXiv:hep-th/0005087].}

\lref\gins{
P. Ginsparg, ``Matrix models of 2d gravity,'' Trieste Lectures (July,
1991), Gava et al., 1991 summer school in H.E.P. and Cosmo.
[arXiv:hep-th/9112013].}

\lref\kostovsix{
I. Kostov, ``Exact Solution of the Six-Vertex Model on a Random
 Lattice,'' Nucl.Phys. B {\bf 575} 513 (2000) [arXiv:hep-th/9911023].}

\lref\hoppe{
J.~Goldstone, unpublished; J.~Hoppe, ``Quantum theory of a massless
relativistic surface,'' MIT PhD thesis, 1982.}

\lref\dvi{R.~Dijkgraaf, C.~Vafa,
``Matrix Models, Topological Strings, and Supersymmetric Gauge
Theories,'' Nucl. Phys. {\bf B644} (2002) 3--20,
[arXiv:hep-th/0206255].}

\lref\dvii{R.~Dijkgraaf, C.~Vafa,
``On Geometry and Matrix Models,'' Nucl. Phys. {\bf B644} (2002)
21--39, [arXiv:hep-th/0207106].}

\lref\dviii{R.~Dijkgraaf, C.~Vafa, ``A Perturbative Window
into Non-Perturbative Physics,'' arXiv:hep-th/0208048.}

\lref\dglvz{
R.~Dijkgraaf, M.T.~Grisaru, C.S.~Lam, C.~Vafa, and D.~Zanon,
``Perturbative Computation of Glueball Superpotentials,''
[arXiv:hep-th/0211017].  }

\lref\cdsw{
F.~Cachazo, M.R.~Douglas, N.~Seiberg, and E.~Witten, ``Chiral Rings
and Anomalies in Supersymmetric Gauge Theory,''JHEP {\bf 0212} (2002) 071
[arXiv:hep-th/0211170].}

\lref\dgkv{
R.~Dijkgraaf, S.~Gukov, V.A.~Kazakov, C.~Vafa, ``Perturbative Analysis
of Gauged Matrix Models,''
Phys.\ Rev.\ D {\bf 68} (2003) 045007 [arXiv:hep-th/0210238].}

\lref\wittft{
E.~Witten, ``Topological quantum field theory,''
Commun. Math. Phys. {\bf 117} (1988) 353.}

\lref\witsdual{
E.~Witten, ``On S duality in Abelian gauge theory,''
Selecta Math.\  {\bf 1} (1995) 383
arXiv:hep-th/9505186}

\lref\mw{
G.~W.~Moore and E.~Witten, ``Integration over the u-plane in Donaldson
theory,'' Adv.\ Theor.\ Math.\ Phys.\ {\bf 1}, 298 (1998)
[arXiv:hep-th/9709193].}

\lref\mamo{
M.~Marino and G.~W.~Moore, ``Integrating over the Coulomb branch in $N
= 2$ gauge theory,'' Nucl.\ Phys.\ Proc.\ Suppl.\ {\bf 68}, 336 (1998)
[arXiv:hep-th/9712062]; ``The Donaldson-Witten function for gauge
groups of rank larger than one,'' Commun.\ Math.\ Phys.\ {\bf 199}, 25
(1998) [arXiv:hep-th/9802185]; ``Donaldson invariants for non-simply
connected manifolds,'' [arXiv:hep-th/9804104].}

\lref\star{
G.~W.~Moore, ``Matrix Models Of 2-D Gravity And Isomonodromic
Deformation,'' lectures at 1990 Cargese Workshop on {\sl Random
Surfaces, Quantum Gravity and Strings,} Prog.\ Theor.\ Phys.\ Suppl.\
{\bf 102}, 255 (1990).  }

\lref\zam{
A.~B.~Zamolodchikov, ``Conformal Scalar Field On The Hyperelliptic
Curve And Critical Ashkin-Teller Multipoint Correlation Functions,''
Nucl.\ Phys.\ B {\bf 285}, 481 (1987).
}

\lref\dfms{
L.~J.~Dixon, D.~Friedan, E.~J.~Martinec and S.~H.~Shenker,
``The Conformal Field Theory Of Orbifolds,''
Nucl.\ Phys.\ B {\bf 282}, 13 (1987).
}

\lref\br{
M.~A.~Bershadsky and A.~O.~Radul, ``Conformal Field Theories With
Additional $Z(N)$ Symmetry,'' Sov.\ J.\ Nucl.\ Phys.\ {\bf 47}, 363
(1988) [Yad.\ Fiz.\ {\bf 47}, 575 (1988)].
}

\lref\lns{
A.~Losev, N.~Nekrasov and S.~L.~Shatashvili,
``Issues in topological gauge theory,''
Nucl.\ Phys.\ B {\bf 534}, 549 (1998)
[arXiv:hep-th/9711108].}

\lref\ackm{
J.~Ambjorn, L.~Chekhov, C.~F.~Kristjansen and Y.~Makeenko,
``Matrix model calculations beyond the spherical limit,''
Nucl.\ Phys.\ B {\bf 404}, 127 (1993)
[Erratum-ibid.\ B {\bf 449}, 681 (1995)]
[arXiv:hep-th/9302014].}

\Title
 {\vbox{
\hbox{hep-th/0211241} 
\hbox{ITFA-2002-53}}}
{\vbox{
\centerline{Matrix Models and Gravitational Corrections}
}}
\bigskip
\centerline{Robbert Dijkgraaf$^{1,2}$, Annamaria Sinkovics$^1$, 
and Mine Tem\"urhan$^1$}
\vskip8mm
\centerline{\sl $^1$Institute for Theoretical Physics}
\centerline{\sl $^2$Korteweg-de Vries Institute for Mathematics}
\centerline{\sl University of Amsterdam}
%\centerline{\sl Plantage Muidergracht 24}
\centerline{\sl 1018 XE Amsterdam, The Netherlands }
\bigskip
\bigskip
\vskip .1in\centerline{\bf Abstract}

\smallskip

We provide evidence of the relation between supersymmetric gauge
theories and matrix models beyond the planar limit.  We compute
gravitational $R^2$ couplings in gauge theories perturbatively, by
summing genus one matrix model diagrams. These diagrams give the
leading $1/N^2$ corrections in the large $N$ limit of the matrix model
and can be related to twist field correlators in a collective
conformal field theory. In the case of softly broken $SU(N)$ $\cN=2$
super Yang-Mills theories, we find that these exact solutions of the
matrix models agree with results obtained by topological field theory
methods.

\vfill

\Date{November, 2002}

%\draft

\newsec{Introduction}

Recently it has become clear that holomorphic or F-term information in
$\cN=1$ supersymmetric gauge theories can be exactly computed using
perturbation theory, when these terms are considered as a function of
the glueball superfield $S$ \dviii\ (for earlier work see among others
\refs{\gv,\vaug,\civ,\ov,\dvi,\dvii}).  Furthermore, for a large class
of theories that allow a large $N$ expansion \`a la 't Hooft \thooft,
the field theory Feynman diagrams, computed in a background super
gauge field, can be seen to reduce directly to the diagrams of a
zero-dimensional bosonic matrix model, where the matrix model
potential is given by the gauge theory tree-level superpotential
\dglvz; for a recent alternative
derivation of this fact using anomalies see \cdsw.

One such holomorphic quantity is the effective superpotential $W_{\rm
eff}(S)$ that is given by summing just the planar diagrams, even for a
finite rank gauge theory. Non-planar diagrams will in general
contribute to gravitational corrections \dviii. In particular,
diagrams with genus one topology, that give the leading $1/N^2$
correction $\cF_1$ to the matrix model free energy, contribute to an
effective curvature term of the form
\eqn\RR{
{1\over 16\pi^2}\int d^4\!x\ \cF_1(S) \, \Tr\,R_+ \wedge R_+,
}
with $R_+$ the self-dual part of the Riemann curvature tensor.  This
induced gravitational correction measures the back-reaction of the
field theory when it is placed in a curved background. The actual
derivation along the lines of \dglvz\ of the matrix model Feynman
rules, directly from the gauge theory Lagrangian in a curved
superspace, will be presented elsewhere
\ref\tobe{ R.~Dijkgraaf, M.T.~Grisaru, C.S.~Lam, C.~Vafa, and
D.~Zanon, ``Planar gravitational corrections for supersymmetric 
gauge theories,'' [arXiv:hep-th/0310061].}.

For exactly solvable matrix models the summation of the diagrams, of
any fixed topology, in closed form can be done in principle, although
the techniques become progressively cumbersome for high genus. One can
thus try to compare these exact answers to known properties of
four-dimensional supersymmetric gauge theories.

In this note we will compare the results for a single matrix model to
the gravitational corrections that have been computed for topological
field theories that are twisted versions of $\cN=2$ supersymmetric
Yang-Mills theories \refs{\witsdual,\mw,\lns,\mamo}. These topological
field theories are used to compute the Donaldson and Seiberg-Witten
invariants of four-manifolds. The gauge theoretic results have been
derived making use of the Seiberg-Witten solution \sw\ and holomorphy
and duality arguments. In this paper we will demonstrate how these
terms can also be computed using loop equations of matrix models.

The genus one correction is simpler for many reasons. Being a first
order correction to a semi-classical evaluation it is given by a
fluctuation determinant, also in the matrix model. If the matrix model
is exactly solvable, this gets often reflected in an emergent geometry,
that in terms of topological string theory arises from a geometrical
transition from an open string to a closed string description. General
arguments tell us that such a dual geometry takes the form of a
non-compact Calabi-Yau three-fold. Topological closed strings
propagating on such a CY three-fold give rise to a genus one 
partition function $\cF_1$ that can be expressed as a generalized
Ray-Singer analytic torsion
\bcov
\eqn\torsion{
\cF_1 = \sum_{p,q=0}^3 p\,q\,(-1)^{p+q} \log\det \D_{p,q},
}
where $\D_{p,q}=\{\dbar,\dbar^\dagger\}$ is the Laplacian acting on
$(p,q)$ forms.  In the simple class of matrix models that we consider
in this paper the effective geometry is essentially given by an affine
algebraic curve, and therefore we expect for general reasons an
expression of the form
$$
\cF_1 =-\hf\log\det \D_0
$$ 
with $\D_0$ the (scalar) Laplacian on the algebraic curve acting on
the collective bosonic field. We will verify this is indeed the case
in some cases by explicit computation. This relation between matrix
models and two-dimensional conformal collective
field theory is a much more general feature, see {\it e.g.} \kii.

The plan of this paper is the following. In section 2 we state the
precise relation between the matrix model and the gauge theory
quantities relevant for gravitational corrections. In section 3 we
consider gravitational couplings in $\cN=2$ super Yang-Mills theories
obtained by a topological twist, and make a comparison to the matrix
model result for $SU(2)$. Then in section 4 we analyze the general
matrix model answer in terms of a collective conformal field
theory. We find that the genus one contribution can be described in
terms of twist field correlation functions with extra dressing to
match the loop equations. This will allow us to make a precise
identification for general $SU(N)$ gauge group.

After this note was finished \ref\kmt{ A.~Klemm, M.~Marino, and
S.~Theisen, `` Gravitational corrections in supersymmetric gauge
theory and matrix models,''JHEP {\bf 0303} (2003) 051 
[arXiv:hep-th/0211216].} appeared that
discusses similar issues.

\newsec{Superpotentials and gravitational couplings}

Let us first briefly review the main results of \refs{\dvi,\dviii} for
the prototypical case of a $U(N)$ gauge theory coupled to single
chiral matter field $\F$ in the adjoint representation. We start with
a tree-level superpotential
\eqn\tree{
\int d^4x d^2\theta\ \Tr\, W(\F),}
where the polynomial $W$ has $n$ distinct critical points. If we
consider a classical vacuum where one distributes $N_i$ of the
eigenvalues of $\F$ in the $i$th critical point of $W$, we have a
classical breaking pattern
$$
U(N) \to U(N_1) \times \cdots \times U(N_n).
$$
The strong coupling dynamics of the corresponding quantum vacuum is
captured by the effective superpotential $W_{\rm eff}(S_i)$ as a
function of the glueball superfields
$$
S_i={1\over 32 \pi^2}\Tr_{SU(N_i)}\cW_\a^2.
$$ 
According to the prescription of \dviii\ this effective superpotential
is given by
\eqn\weff{
W_{\rm eff}(S) =\sum_i \left[
N_i {\d \cF_0 \over \d S_i} + 2\pi \tau_0\, S_i\right],
}
where $\tau_0$ is the bare coupling and $\cF_0(S_i)$ is the free
energy of the corresponding matrix model, obtained in a semi-classical
expansion around the classical vacuum.

This matrix model takes the form of an integral over a $\tN \times
\tN$ matrix $\F$ (here we carefully distinguish between $\tN$, the rank of 
the matrix model, and $N$, the rank of the gauge theory)
\eqn\matrixmodel{
{1\over {\rm vol}\,U(\tN)}\int d\F\,\exp\[-{1\over g_s}\Tr\,W(\F)\] =
\exp \[-\sum_{g\geq 0} g_s^{2g-2} \cF_g(S_i)\], }
with the identification $S_i=g_s \tN_i$ in the 't Hooft limit $g_s\to
0,$ $\tN_i \to \infty$. More precisely, we have
\eqn\fzero{
\cF_0(S) = \half S^2 \log (S/\L_0^3) + \cF_0^{\rm pert}(S).
}
The first term gives rise to the Veneziano-Yankielowicz
effective action of the pure Yang-Mills theory \vy,
\eqn\vyterm{
W_{\rm eff}(S) = N S \log(S/\L^3).
}
In the matrix model this contribution to \fzero\ is reproduced as the
large $\tN$ volume of the unitary group \ov.

The second term in \fzero\ is given by a sum over planar diagrams that
appear in the perturbative expansion of the matrix model. (See \dgkv\
for a careful description of this expansion around a vacuum with a
spontaneous broken gauge symmetry.) A diagram with $\ell$ index loops
comes with a factor of $S^\ell$. The actual physical values of $W_{\rm
eff}$ and the condensates $S_i$ in the quantum vacua are given by
extremizing \weff\ with respect to the glueball fields $S_i$.

As we mentioned in the introduction this relation is not restricted to
planar diagrams. There is an elegant interpretation of the higher
genus diagrams that give the corrections $\cF_g$'s in terms of the
coupling to a supergravity background \dviii. In particular, the
induced gravitational effective action obtained by putting the field
theory on a curved space-time contains the F-term
\eqn\grav{
{1\over 16\pi^2} \int d^4\!x\ \cF_1(S) \, \Tr\,R_+ \wedge R_+.  }
where $R_+$ is the self-dual part of the Riemann tensor. (There is of
course a similar anti-holomorphic term $\overline{\cF}_1$ multiplying
$\Tr\,R_- \wedge R_-$.)

If we consider the partition function on a Euclidean four-manifold
$M^4$, then this gravitational coupling induces a term
\eqn\tft{
\exp \cF_1(S)\({1\over 2}\chi -{3\over 4}\sigma\),
}
with $\chi$ the Euler number and $\sigma$ the Hirzebruch signature of
$M$. 

\ifig\nonplanar{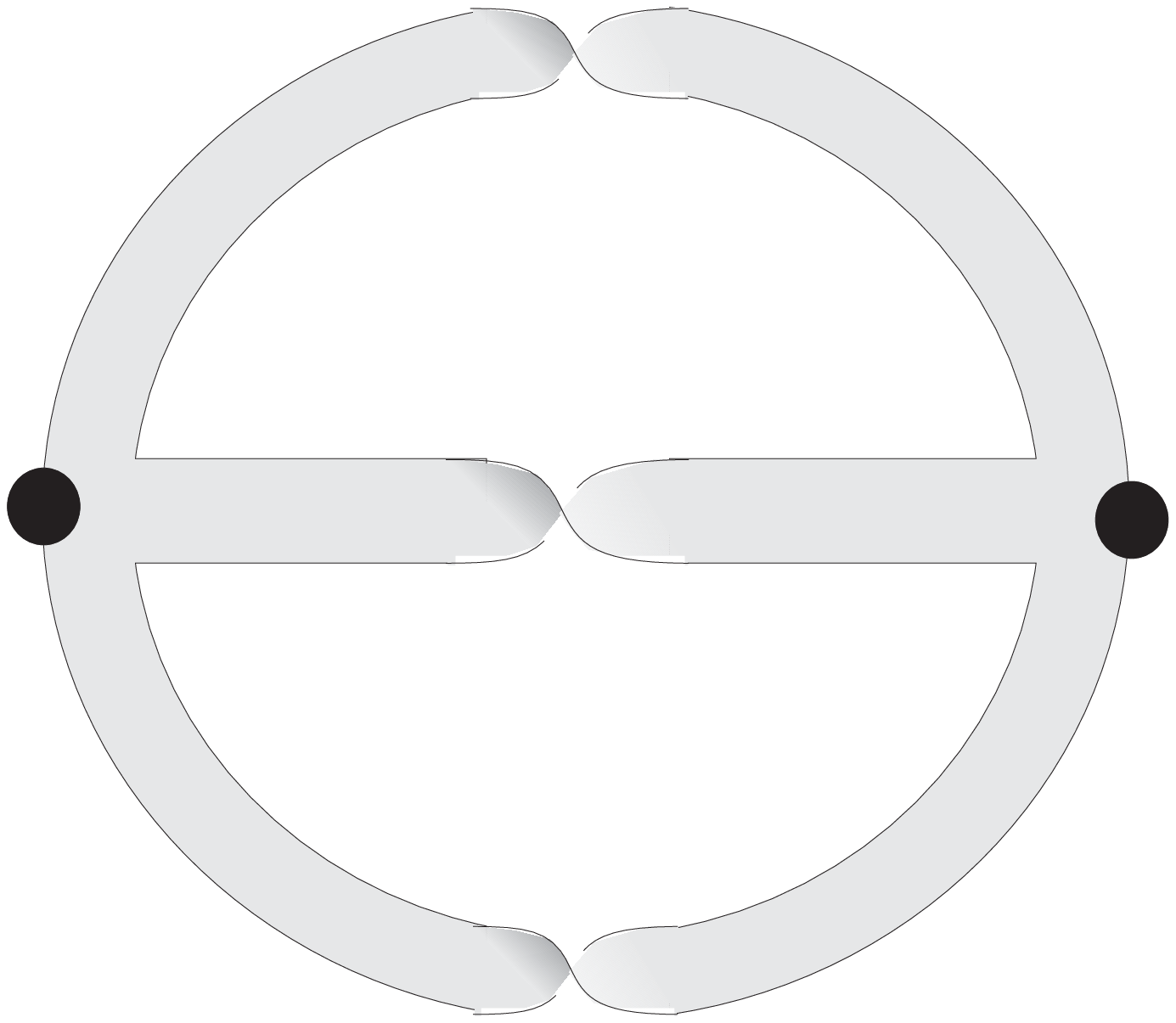}{40}{
The simplest genus one, non-planar diagram in a cubic theory --- the
leading perturbative contribution to $\cF_1$. The $\bullet$'s indicate
insertions of the background gauge field $\cW_\a$.}

Evaluating the term $\cF_1(S)$ in perturbation theory, one finds that
it is given exactly by the sum of diagrams with topology genus one,
{\it i.e.}\ the diagrams that give the leading $1/\tN^2$ corrections
in the large $\tN$ limit of the matrix model. More precisely,
$\cF_1(S)$ is given as
\eqn\fone{
\cF_1(S) = -{1\over 12} \sum_i \log (S_i/\L_0^3) + \cF_1^{\rm pert}(S).
}
This expression is the gravitational analogue of \fzero. Assuming confinement,
so that the only the field accounted for is $S$, the first term has an 
interpretation as an integrated form of the gravitational contribution 
to the $U(1)$ R-anomaly, 
\eqn\anom{
\d_\m J_5^\m = {1\over 16 \pi^2}\left[{1\over 2}
\Tr_{U(N_i)} F\wedge F - {1\over 12} 
\Tr\, R \wedge R\right].
}

If one assumes that the low energy dynamics of the gauge system is
described by an effective action in which the glueball superfields
$S_i$ can be treated as elementary fields, the anomalous behaviour
under the R-symmetry
$$
S \to e^{i\theta} S
$$
is reproduced by the combination of the Veneziano-Yankielowicz
contribution $NS\log S$ to $W_{\rm eff}(S)$ --- recall that the top
component of $S$ is $\Tr\,F_+\wedge F_+$ --- together with the $-{1\over
12}\log S$ multiplying the gravitational correction. Including the
complex conjugated term that muliplies $\log \overline{S}$, we see
that we pick up precisely the anomaly \anom.

The perturbative contribution $\cF_1^{\rm pert}$ to \fone\ is given by
summing all genus one diagrams. For example, in a cubic theory, with
superpotential $W(\F)=m\F^2 + g \F^3$ the leading diagram is given by
\nonplanar\ and this gives
$$
\cF_1^{\rm pert}(S) = \hf {g^2\over m^3} S + \cO(S^2)
$$

Of course in the physical vacua all these expressions for $\cF_1$ have
to be evaluated for those values of the $S_i$ that minimize the
effective superpotential given by the planar contribution.
\foot{
The genus zero diagrams can also contribute, but their contribution can
be shown to cancel at the critical point.}

\newsec{Matrix models and $\cN=2$ theories}

Matrix model methods can be used in particular to find the celebrated
solution to the pure $\cN=2$ super Yang-Mills theory of Seiberg and
Witten \sw. To describe the $SU(N)$ gauge theory one breaks the
supersymmetry down to $\cN=1$ by introducing a degree $N+1$ tree-level
superpotential $\Tr\,W(\F)$ of the adjoint chiral multiplet $\F$ and
picks the breaking pattern
$$
U(N) \to U(1)^N,
$$
by distributing the $N$ eigenvalues of $\F$ equally among the $N$
critical points of $W$; that is, one chooses all $N_i=1$. One further
decouples the diagonal $U(1)$ by putting the overall bare coupling
$\tau_0=0$. The effective superpotential then simplifies to
$$
W_{\rm eff}(S) = \sum_i {\d \cF_0 \over \d S_i}.
$$

In this case the planar diagrams can be exactly summed  and the
solution can be written in terms of period integrals on the associated
hyperelliptic Riemann surface \dvi
$$
y^2 = P(x)^2+ f(x),\qquad P(x) = W'(x)=\sum_{i=0}^N u_i x^{N-i}.
$$
In this case the definition of the variables $S_1,\ldots,S_N$ is
subtle, since they are defined in terms of the traceless piece of a
$U(1)$ gauge field. Classically they vanish but as operators they make
sense quantum mechanically \cdsw. The dependence on the $S_i$ is
implicit in terms of the quantum deformation $f(x)$, a polynomial of
degree $N-1$. After solving the constraint $d W_{\rm eff}(S)=0$, this
curve takes the familiar SW form
\eqn\swcurve{
y^2 = P(x)^2 -\L^{2N}.
}
By the introduction of the bare superpotential we have effectively
localized to a particular point of the Coulomb branch of the $\cN=2$
theory.

To obtain the original $\cN=2$ model one can now scale the tree-level
superpotential as $W \to \e\, W$ and take the limit $\e \to 0$. There
are two obvious quantities that by a scaling argument do not depend on
$\e$ and can therefore be straightforwardly extracted from the $\cN=1$
solution\foot{For the cubic potential $W=m\F^2 + g\F^3$ this can be
explicitly checked by scaling $m\to \e m,$ $g\to \e g$, $S \to \e
S$. Note that in the matrix model $S=g_sN$ and therefore $g_s \to \e
g_s$.}. First, there is the coupling matrix
$$
\tau_{ij} = {\d^2 \cF_0 \over \d S_i \d S_j}
$$
of the $U(1)^N$ low-energy effective Abelian theory. Geometrically
this is given by the period matrix of the curve \swcurve. The second
$\e$-invariant quantity is the genus one free energy $\cF_1$ that
gives the gravitational correction \grav.

\subsec{Gravitational coupling from topological field theory}

On flat spacetime the ${\cal N}=2$ $SU(2)$ gauge theory is described
by the Seiberg-Witten solution. Putting the theory on a curved
manifold additional gravitational terms appear in the low energy
effective action.  This gravitational correction has been directly
computed in the $\cN=2$ theory --- more precisely, in a topological
twisted version of the theory that computes Donaldson invariants. In
the twisted version one modifies the action of the Lorentz group
$$
SO(4) \cong SU(2)_+ \times SU(2)_-.
$$ 
One replaces $SU(2)_+$ with the diagonal subgroup of $SU(2)_+ \times
SU(2)_R$, where the last factor is the $\cN=2$ internal R-symmetry
group \wittft.

In the twisted topological theory considered on a curved four-manifold
$M$ these interactions are restricted to the topological $R \wedge
R^*$ and $R \wedge R$ terms (with $R_\pm=\hf(R\pm R^*)$) 
proportional to the Euler number
$\chi(M)$ and the Hirzebruch signature $\s(M)$ respectively.  The
gravitational couplings contribute to the partition function with the
factor
\eqn\partg{
\exp\[b(u) \chi  + c(u) \sigma\], 
}
where $b(u)$ and $c(u)$ are functions of the parameter $u$ on the
gauge theory moduli space.

The precise form of the functions $b(u),c(u)$ can be inferred from
analyzing the modular transformation properties of the quantum theory
on the curved manifold. Cancellation of the modular anomaly and
additional input from the singularity structure of Seiberg-Witten
moduli space determines the measure contribution
\refs{\witsdual,\mw,\lns,\mamo}.

To connect these computations in topological field theory to the
physical theory we recall that for manifolds with metrics of $SU(2)$
holonomy (hyper-K\"ahler manifolds) the topological twist is invisible
since there is no holonomy in $SU(2)_+$. We can therefore directly
compare to the physical gauge theory. In that case the metric is pure
self-dual, and we have
$$
\sigma = -{2\over 3}\chi.
$$
For example one could take $M=K3$ for which $\chi=24$ and $\sigma=-16$.
So the overall contribution to the path-integral is
$$
\exp\[\(b(u) -{2\over3}c(u)\)\chi\].
$$
If we compare this to \tft, where we use that for a self-dual geometry
${1\over 2} \chi -{3\over 4} \sigma=\chi$, we have the following
identification between the matrix model and gauge theory quantities
$$
\cF_1(S) = b(u) - {2\over 3}c(u).
$$
We will now check this relation in a number of cases. We will for
convenience put $\chi=1$.

\subsec{The $\cN=2$ $SU(2)$ theory }

In this case the Seiberg-Witten geometry can be described by deforming
the $\cN=2$ theory with a tree level superpotential,
\eqn\pot{
W'(\F) = \eps ( \F^2 - u).
 } 
(For more details about the perturbative derivation of this particular
case see \dgkv.) As described above, extremization of the effective
glueball superpotential gives the Seiberg-Witten curve for $SU(2)$
\eqn\curve{
y^2 = (x^2 - u)^2 - 1.}
Here the scale $\L$ is set to one for convenience and the factor $\e$
is absorbed.  As we mentioned the physical quantities $\cF_1$ and the
coupling matrix $\t_{ij}$ are independent of the deformation parameter
$\eps$. The curve has four branch points at
\eqn\brpoint{ x_i = \pm \sqrt{ u \pm 1}. } 
It is described by the two-cut solution of the matrix model with the
potential $W(\F)$ given by \pot.

The genus one free energy for two-cut solutions in matrix models have
been explicitly computed. Here we use the relevant solution of Akemann
\akemann, which is an elaboration of the methods of \ackm,
\eqn\Fakemann{
\eqalign{
\cF_1 = &  - {1 \over 24} \sum_{i=1}^4 \log{M_i}  
- \hf \log{|K(k)|} \cr & \qquad - {1\over 12} \D + {1 \over 4}
\log{|(x_1 - x_3)(x_2 - x_4)|}.
\cr} 
}
This solution was derived by an iterative genus expansion of the loop
equation; we discuss this further in the next section. Here $\D$ is
the discriminant of the elliptic curve \curve\
\eqn\discr{ 
\D=\prod_{i<j} (x_i - x_j)^2 =64(u^2-1),
}
and $K(k)$ is the complete elliptic integral, where the nome $k$ is
expressed in the modulus $\tau$ of the SW curve. The solution also
depends on the first moments of the potential that are generally defined as
\eqn\moment{
M_i = {1 \over 2 \pi i} \oint_{C_\infty} dx {W'(x) \over (x - x_i) 
\sqrt{ \prod_{i=1}^4 (x - x_i)}}.
}
For the simple potential \pot\ the contour can be deformed to
infinity, and one gets $M_i=\eps$.

For comparison with the gauge theory result it is useful to express
$\cF_1$ in terms of the parameter of $SU(2)$ moduli space $u$. The
elliptic parametrization of the SW curve \curve\ can be written in
terms of Jacobi $\th$ functions as
$$
u = { \th_2^4 + \th_3^4 \over 2 (\th_2 \th_3)^2},
\qquad
u^2 - 1 = {\th_4^8 \over 4 (\th_2 \th_3)^4 } , 
$$ 
where the definition of the $\th$ functions is as usual
$$
\eqalign{
\th_2 & = \sum_{n \in \Z} q^{\hf ( n + \hf)^2} \cr
\th_3 &= \sum_{n \in \Z} q^{\hf n^2} \cr
\th_4 & = \sum_{b \in \Z} (-1)^n q^{\hf n^2}} 
$$
with $q= e^{2 \pi i \t}$. A useful identity they satisfy is $\th_2^4 +
\th_4^4 = \th_3^4$. The complete elliptic integral $K(k)$ can also be
expressed in $\th$ functions as
$$
K(k) = {\pi \over 2} \th_3^2.
$$ 
With these elliptic parametrization the matrix model answer for the
two-cut solution to $\cF_1$ can be written as
\eqn\fff{
\cF_1 = - {1 \over 6} \log{\eps} + 
{1 \over 4} \log{ {4 \over \pi (\th_2 \th_3)^2} } 
- {1 \over 12} \log{ {16\, \th_4^8 \over (\th_2 \th_3)^4} }. 
} 
The factor $\log \e$ can be absorbed in the measure.

\subsec{Comparison to the gauge theory}

We now have to compare this result to the topological field theory
answer  that reads \refs{\witsdual,\mw}
\eqn\partg{
\eqalign{
e^{b(u)} & = \a \left( (u^2 -1) {d \t \over d u} \right)^{1/4} ,\cr
e^{c(u)} & = \b (u^2 -1)^{1/8},\cr
}}
where $\a$ and $\b$ are constant coefficients.  This contribution to
the partition function should match with the matrix theory computation
for the corresponding genus one contribution.  To check this, it helps
to rewrite the gauge theory contribution as
\eqn\gauge{
Z_{\rm gauge} = e^{b(u)-{2\over 3} c(u)}=A^{-1/2} \D^{-1/12},
}
with 
$$
A = {d a \over d u},\qquad \D = 64(u^2 - 1).
$$
Here 
$$
(u^2 -1) {d \t \over d u} = {i \over 4 \pi}
\left( {d u \over d a} \right)^2
$$
is rewritten in terms of the ``electric'' period of the Seiberg-Witten
curve $a$.  Substituting the modular parametrization of the curve in
terms of $\th$ functions we find
$$
\Delta = {16 \th_4^8 \over  (\th_2 \th_3)^4 }, \qquad
A = {da \over du} = \hf \th_2 \th_3,
$$ 
%we find
%%
%\eqn\Zf
%{Z_{\rm gauge} = (\th_4)^{-1/6} {\eta}^{-5/6} }
%
Comparing with the matrix theory contribution \fff\ we find perfect
agreement.

\subsec{$SU(N)$ generalization}

The gauge theory computation for the partition function can be
generalized for the $SU(N)$ theory. The generalization is based on a
similar analysis of anomalies as for the $SU(2)$ case.

At a generic point on the Coulomb branch, where the gauge symmetry is
broken to $U(1)^{N-1}$, the $SU(N)$ theory can be described by the
hyperelliptic curve
$$
\eqalign{
y^2 &= P(x)^2 - 1 = \prod_{i=1}^{2 N } (x -x_i), \cr
P(x) & = \sum_{i=0}^N u_i x^{N-i}, \quad \sum_{i=1}^N u_i =0.}
$$
Here the $u_i$'s are the symmetric polynomials of the roots of $P(x)$,
and $x_i$ are the branch points of the curve.  The hyperelliptic curve
is a Riemann surface of genus $g=N-1$. 

For a genus $g$ Riemann surface one takes a basis of $2 g$ homology
cycles $(A_i, B_i)$ with canonical intersection product. The periods
of the curve are then related to a set of dual holomorphic one-forms
$\w_i=x^{i-1} d x/y$ as (we choose the homology basis slightly
different then in \refs{\lns,\mamo} in order to make contact with the matrix
model basis)
$$ 
A_{ij} = \oint_{A_i} \w_j = {\d a_{i} \over \d u_{j+1}},
\qquad B_{ij} =  \oint_{B_i} \w_j =  {\d a_{D,i} \over \d u_{j+1}}.
$$
The period (or coupling) matrix $\t_{ij}$ is given as
$$
\t_{ij} = {\d a_{D,i} \over \d a_j} = (B A^{-1})_{ij}.
$$ 

The partition function for $SU(N)$ is a direct generalization of the
corresponding $SU(2)$ contribution \refs{\lns,\mamo}. We will write it
as (discarding overall constants)
\eqn\zsuN{
Z_{\rm gauge}= A^{-\chi/2} \D^{\s/8},
}
with
$$
A = \det A_{ij},  \qquad
\Delta = \prod_{i<j}^{2N}  (x_i - x_j)^2.
$$
Putting a self-dual metric and $\chi=1$ we get
\eqn\sun{
Z_{\rm gauge}= A^{-1/2} \D^{-1/12}.
}
To compare this result to the genus one free energy of the matrix
model we first have to explain how matrix model results can be
computed using conformal field theory.
 
\newsec{Multicut solutions and conformal field theory}

\subsec{Loop equations and Virasoro constraints}

For the one-loop free-energy for the $SU(N)$ theory we have to solve
the corresponding matrix model with the tree-level superpotential
$W(\F)$ with $ W'(\Phi) = \epsilon P(\Phi)$ and with the maximum
number of cuts. The most efficient way to derive multicut solutions
for matrix models is by using loop equations and conformal field
theory techniques. The method of using loop equations to obtain the
$1/N$ corrections in matrix models was developed in \ackm. Here we
will follow closely \kii\ that gives a good general exposition of
the relation of these methods to conformal field theory.

We start from the partition function of the associated matrix model
$$
Z = {1\over {\rm vol}\,U(\tN)}
\int d \Phi \ \exp\[- {1\over g_s} \Tr\, W(\Phi)\],
$$ 
for the $\tN \times \tN$ matrix $\F$ with a general potential
$W(\Phi)$.  The reparametrization invariance of this integral leads
directly to the so-called loop equation\foot{These loop equations have
been recently given a gauge theoretic interpretation in
\cdsw.}. 
The simplest way to derive the loop equation is taking a shift $\Phi
\rightarrow {\delta \over (x -
\Phi)}$ where $\delta$ is a small number. This gives the equation for
the loop correlator
$$
\left\langle \w(x)^2 - {1\over g_s}
 \Tr\left({ W'(\Phi) \over {x- \Phi}}\right) 
\right \rangle  =0,
$$ 
where
$$
\w(z) = \Tr {1 \over x- \Phi} = \sum_{I=1}^\tN {1 \over x - \l_I}
$$
is the loop operator and $\l_I$  the eigenvalues of
the matrix $\F$. For a general potential with coupling constants $t_n$
$$ 
W(\Phi) = -\sum_{n=1}^{\infty} t_n \Phi^n,
$$
the loop equation can be rewritten as
$$ 
\oint {d x' \over 2 \pi i} { 1 \over x -x'} \left\langle T(x') 
\right\rangle =0,
$$
where the contour includes all eigenvalues $\l_I$ but excludes the
point $x$. Here we introduce the stress-tensor $T(x)$ of the
collective field $\v(x)$
$$
\eqalign{
T(x) &= \hf (\d \v(x))^2, \cr
\v(x) &=  W(x) -2 g_s \,  \Tr\, \log \left( {1 \over 
x-\F} \right). }
$$ 
The loop equation can be reformulated as the Virasoro constraints
\refs{\dvv,\kawai}
$$
\eqalign{ 
L_n Z &=0, \quad n \ge -1, \cr L_n &= \sum_{k=0}^n {\d \over \d t_k}
{\d \over
\d t_{n -k}} + \sum_{k=0}^{\infty} k t_k {\d \over \d t_{n+k}}},
$$
where the operators $L_n$ satisfy (half of) the Virasoro algebra
$$
[L_n, L_m] = (m-n) L_{m+n}.
$$
The loop equation can be solved iteratively order by order making a
${1/ \tN}$ expansion. The planar limit is usually not so hard to
solve, but for the next order solution one needs special techniques.

\subsec{Planar solution and effective geometry}

To solve the loop equation in the $\tN \rightarrow \infty$ limit, it
is simplest to rewrite it in terms of the polynomial
$$ 
f(x) = 4 g_s \left\langle \Tr {W'(\Phi) - W'(x) \over
\Phi -x } \right\rangle
$$ 
If $W(x)$ is of degree $n+1$ then the polynomial $f(x)$ is of degree
$n-1$. Denoting the classical average of the loop operator as
$$ 
\w_c(x) = {1 \over \tN} \langle \w(x) \rangle,
$$
in the large $\tN$ limit the loop equation becomes quadratic
$$
\w_c(x)^2 -{1\over g_s \tN} W'(x) \w_c(x) + {1\over 4g_s^2 \tN^2} 
f(x) =0.
$$
In terms of the bosonic collective field $\v(x)$ we can say that this
obtains a large vacuum expectation value in the planar limit, which is
the solution for the classical Virasoro constraint.  The current
$\d\v(x)$ takes its classical value
$$
\eqalign{
\d\v_c(x) & = W'(x) - 2 g_s \tN  \w_c(x) \cr 
          & = W'(x) - {2 g_s \tN \over x} + 
\cO\left({1 \over x^2} \right).\cr
}
$$
For our solution we have with $y=\d\v_c(x)$
$$
y^2 = W'(x)^2 + f(x).
$$
We can write this as the hyperelliptic curve
\eqn\hyperell{
y^2=\prod_{i=1}^{2N}(x-x_i).  
} 
Since the classical field $\v(x)$ changes sign around the branch
points, the expectation values of the bosonic field are given by two
branches of $y(x)$.  Then $\v(x)$ can be thought of as a single
bosonic field defined on the branched covering given by the
hyperelliptic curve \hyperell. The classical value of the collective
field is given as
\eqn\class{
\d\v_c(x) = \prod_i (x-x_i)^{1/2}.
}

\subsec{Subleading corrections and twist fields}

The subleading term $\cF_1$ in the free energy is given by the
Gaussian fluctuations around the classical solution.  As we mentioned
in the introduction, by general arguments this term is equal to $- \hf
\log {\rm det} \Delta_{0}$, the Laplace operator on the Riemann surface,
and additional (dressing) terms arising from the fluctuation of the
branch points.

Instead of thinking of $\v(x)$ as a field living on the hyperelliptic
Riemann surface \hyperell, we can also think of it on the complex
$x$-plane in the presence of twist operators $\s(x_i)$ associated with
the branch points $x_i$. In the neighbourhood of such a twist field
the current $\d\v(x)$ is no longer single-valued but has a branch cut
in its operator product
$$
\d\v(x) \cdot \s(x_i) \sim (x-x_i)^{-1/2} \tau(x_i).
$$
Here $\s(x)$ and $\tau(x)$ are conformal fields of dimension $1/16$ and
$9/16$ respectively.
A naive expression for the genus one contribution to the matrix model
would now be given by
$$
Z_{\rm twist} = \Bigl\langle\ \prod_{i=1}^{2N} \s(x_i)\, \Bigr\rangle.
$$
The chiral\foot{In this expression the chiral projection is done by
putting the loop momenta of the field $\v(x)$ to zero.} twist field
correlation function is well-known \refs{\dfms,\zam,\br}
\eqn\ztaist{
Z_{\rm twist} =  A^{-\hf}  \prod_{i <j} (x_i - x_j)^{-1/8}. 
}
Here 
$$
A = \det (A_{ij})
$$
is the determinant of the period matrix, related to the integral of
the one-forms $\w_i$ over the $A$-cycles
\eqn\matrixA{  
A_{ij} = \oint_{A_i} {x^{j-1} d x \over y}.
}

For example, in the case of a two-point function we get the familiar
result
$$
\bigl\langle \s(x_1)\s(x_2) \bigr\rangle = (x_1-x_2)^{-1/8}
$$ 
expressing the fact that the conformal dimension of a $\Z_2$ twist
field is $1/16$. 

Formula \ztaist\ also can be expressed as the chiral determinant of
the Laplace operator $\D_0$ of the twisted boson on the hyperelliptic
curve $$ Z_{\rm twist} = \(\det\,\D_0\)^{-1/2}.  $$

\subsec{Star operators}

However, \ztaist\ is not the full answer, since as it stands this
expression does not solve the Virasoro constraints. An elegant
solution to this has been given by Kostov in terms of star operators
\kii.

We can associate a Hilbert space with the local complex variable near
each branch point and solve the Virasoro constraint in the vicinity of
the branch point. We have to look for an operator which creates a
conformally invariant state near the branch point. The twist operator
itself does not satisfy all the Virasoro constraints, in particular it
does not satisfy $L_{-1}$.  Therefore we will look for a new operator
which satisfies all constraints.  Such operators are called star
operators \star, and they are constructed from the modes of the
twisted bosonic field near the branch point\foot{We would like to
thank I.~Kostov for sharing with us some unpublished work on the
construction of the star operators.}.

The twisted bosonic current near the branch point $x_i$ is now
decomposed into a classical and quantum part
$$ 
\d \v(x) = \d \v_c (x) + \sum_{r \in {\bf Z} + \hf} \a_r 
(x - x_i)^{-r -1}.
$$
The expansion of the classical current \class\ is
$$ 
\d \v_c(x) = \sum_{r \ge \hf} \mu_r(x_k) \cdot (x - x_k)^{r - 1}. 
$$
This defines the coefficients $\mu_r(x_i)$.  The Fock vacuum for such
a twist field is defined as
$$
|0_i \rangle = \sigma(x_i) |0 \rangle,
$$
and it satisfies
$$
\alpha_r |0_i \rangle =0,\quad r>0. 
$$
Since it depends on the position of the branch points it is not
translationally invariant. To make it invariant, one introduces the
star operator
$$ 
S(x_i) = e^{s(x_i)} \s(x_i) 
$$ 
and assume it is defined perturbatively by a mode expansion
$$ 
s(x_i) = \sum_{n \ge 0} {1 \over n!} \sum_{r_1 \ldots r_n} s_{r_1
\ldots r_n}(x_i) \a_{-r_1} \ldots \a_{-r_n}.
 $$
The coefficients in the mode expansion are determined by imposing the
conditions of conformal invariance
$$ 
L_n \, e^{s(x_i)} |0_i \rangle =0, \quad n \ge -1. 
$$
Up to $1/\tN^2$ correction one finds simply an extra multiplicative factor
\kii
$$
S(x_i) = \[\mu_{3/2}(x_i)\]^{-{1/ 24}} \s(x_i).
$$

The full genus one contribution to the free energy, obtained by
solving the loop equation including the order $1/\tN^2$ corrections,
is therefore given by the correlation function of star operators, not
the twist operators,
\eqn\twists{
\eqalign{
{\cal F}_1 & = \log \Bigl\langle\ \prod_{i=1}^{2N} S(a_i)\,
           \Bigr\rangle \cr & = - {1 \over 24} \sum_{i=1}^{2 N} \log
           \mu_{3/2}(x_i) +
\log Z_{\rm twist},
}} 
where $Z_{\rm twist}$ is the correlation function of the $2N$ twist
fields \ztaist.  From the expansion of the classical current $\d
\v_c(z)$ we get
$$ 
- {1 \over 24} \sum_{i=1}^{2N} \log \mu_{3/2}(x_i) = - {1 \over 24}
\log \prod_{i<j} (x_i - x_j)
$$
So the final result for $Z_{\rm matrix}$ is then\foot{Note that this
form is given incorrectly in \kii.}
\eqn\fmatrix{
Z_{\rm matrix} = e^{\cF_1} = A^{-1/2} \Delta^{-1/12}
}
This result is in complete agreement with the gauge theory partition
function $Z_{\rm gauge}$ \sun.

\bigskip
\centerline{\bf Acknowledgements}

We would like to thank I.~Kostov and C.~Vafa for enlightening
discussions. This research is supported by NWO, the FOM Programme {\it
String Theory and Quantum Gravity}, and the CMPA grant of the
University of Amsterdam.

\listrefs

\bye